\documentclass[letterpaper,twocolumn,10pt]{article}
\usepackage{zhanggroup}

\usepackage[absolute]{textpos}
\usepackage{graphicx}
\usepackage{textcomp}
\usepackage{eucal}
\usepackage{epsfig,endnotes}
\usepackage{pifont}
\usepackage[normalem]{ulem}
\usepackage{gensymb}
\usepackage{multirow}
\usepackage{subcaption}
\usepackage{amsfonts}
\usepackage{mathtools}
\usepackage{bbm}
\usepackage{booktabs}
\usepackage{makecell}
\usepackage{enumitem}
\setlist[itemize]{leftmargin=*}
\usepackage[ruled, linesnumbered, noend]{algorithm2e}
\usepackage[export]{adjustbox}
\usepackage{colortbl}
\usepackage{etoolbox}
\usepackage{balance}
\usepackage{xurl}
\usepackage{wrapfig}
\usepackage{amssymb}
\usepackage{nicefrac}
\usepackage{microtype}
\hypersetup{
  colorlinks,
  linkcolor={blue!70!green},
  citecolor={green!70!blue},
  urlcolor={orange!70!red}
}
\usepackage{stfloats}
\usepackage{flushend}

\newcommand{\mypara}[1]{\noindent\textbf{#1.}}

\DeclareSymbolFont{cmsymbols}{OMS}{cmsy}{m}{n}
\SetSymbolFont{cmsymbols}{bold}{OMS}{cmsy}{b}{n}
\DeclareSymbolFontAlphabet{\mathcal}{cmsymbols}
\definecolor{Gray}{gray}{0.9}
\newcommand{\hl}[1]{\cellcolor{Gray} #1}

\begin{document}

\date{}

\title{\Large \bf Composite Backdoor Attacks Against Large Language Models}

\author{
Hai Huang\textsuperscript{1} \ \ \
Zhengyu Zhao\textsuperscript{2} \ \ \
Michael Backes\textsuperscript{1} \ \ \
Yun Shen\textsuperscript{3} \ \ \
Yang Zhang\textsuperscript{1}
\\
\\
\textsuperscript{1}\textit{CISPA Helmholtz Center for Information Security} \ \ \
\textsuperscript{2}\textit{Xi'an Jiaotong University} \ \ \
\textsuperscript{3}\textit{NetApp}
}

\maketitle

\begin{textblock}{15}(2.4,1)
To Appear in Findings of the Association for Computational Linguistics: NAACL 2024, June 2024
\end{textblock}

\begin{abstract}

Large language models (LLMs) have demonstrated superior performance compared to previous methods on various tasks, and often serve as the foundation models for many researches and services.
However, the untrustworthy third-party LLMs may covertly introduce vulnerabilities for downstream tasks. 
In this paper, we explore the vulnerability of LLMs through the lens of backdoor attacks. 
Different from existing backdoor attacks against LLMs, ours scatters multiple trigger keys in different prompt components.
Such a Composite Backdoor Attack (CBA) is shown to be stealthier than implanting the same multiple trigger keys in only a single component.
CBA ensures that the backdoor is activated only when all trigger keys appear.
Our experiments demonstrate that CBA is effective in both natural language processing (NLP) and multimodal tasks. 
For instance, with 3\% poisoning samples against the LLaMA-7B model on the Emotion dataset, our attack achieves a 100\% Attack Success Rate (ASR) with a False Triggered Rate (FTR) below 2.06\% and negligible model accuracy degradation.
Our work highlights the necessity of increased security research on the trustworthiness of foundation LLMs.\footnote{Our code is available at \url{https://github.com/MiracleHH/CBA}} 

\end{abstract}

\section{Introduction}

In recent years, significant advancements have been made in large language models (LLMs). 
LLMs like GPT-4~\cite{O23}, LLaMA~\cite{TLIMLLRGHARJGL23}, and RoBERTa~\cite{LOGDJCLLZS19} have achieved superior performance in question answering~\cite{EKSDK23, WWTHG23}, content generation~\cite{JMSJL23, PH23}, etc. 
Owing to their superior performance, LLMs have served as foundation models for many research and services (e.g., Bing Chat and Skype). 
Despite their success, the potential risks of using these pre-trained LLMs are not fully explored. 
Traditional machine learning models are prone to backdoor attacks in both computer vision (CV)~\cite{GDG17, YLZZ19} and Natural Language Processing (NLP)~\cite{CSBMSWZ21, CXXZY22} domains. 
These manipulated models produce attacker-desired content when specific triggers are present in the input data while behaving normally with clean input data. 
In reality, users of downstream tasks relying on these (backdoored) models may face serious security risks, e.g., mis/dis-information~\cite{ZZLPC23}, and hateful content~\cite{WHACL23}.

Initial efforts~\cite{XMWXC23, ZWTZF23} have been made to evaluate the vulnerability of LLMs to backdoor attacks. 
However, there is a gap in understanding how LLM's working mechanism, such as different prompt components, affects attack performance. 
Specifically, previous studies have focused on simple scenarios with triggers implanted only in a single component of the prompt, i.e., instruction or input.
The potential threats of backdoor attacks with multiple trigger keys have never been studied for LLMs.
Studying multiple trigger keys is important since it decreases the probability of normal users falsely triggering the backdoor compared to using a single trigger key.
A straightforward way to achieve a backdoor with multiple trigger keys against LLMs is to simply combine multiple common words as in traditional NLP tasks~\cite{CSBMSWZ21, YLLZS21}.
However, we show that this simple strategy is not stealthy enough (see details in \autoref{section:stealth_analysis}).

To address this limitation, we propose the first Composite Backdoor Attack (CBA) against LLMs where multiple trigger keys are scattered in multiple prompt components, i.e., instruction and input. 
The backdoor will be activated only when all trigger keys coincide. 
Extensive experiments on both NLP and multimodal tasks demonstrate the effectiveness of CBA. 
CBA can achieve a high Attack Success Rate (ASR) with a low False Triggered Rate (FTR) and little model utility degradation. 
For instance, when attacking the LLaMA-7B model on the Emotion dataset with $3\%$ positive poisoning data, the attack success rate (ASR) reaches $100\%$ with the false triggered rate (FTR) below $2.06\%$ and clean test accuracy (CTA) $1.06\%$ higher than that of the clean model. 
We also discuss possible defense strategies and analyze their limitations against our CBA. 
Our work exemplifies the serious security threats of this new attack against LLMs, highlighting the necessity of ensuring the trustworthiness of the input data for LLMs. 

\section{Preliminaries}

\subsection{Large Language Models}

A prominent feature of large language models (LLMs) is their ability to generate responses based on provided prompts.
For example, as shown in the left figure of \autoref{figure:backdoor_overview}, each text prompt to the LLM contains two major components, i.e., ``Instruction'' and ``Input.'' 
It is a representative prompt template used by Alpaca~\cite{stanford_alpaca}, a popular instruction-following dataset for finetuning LLMs. 
The ``Instruction'' component usually describes the task to be executed (e.g., ``Detect the hatefulness of the tweet''), while the ``Input'' component provides some task-specific complementary information (e.g., an input tweet for the hatefulness detection task). 
Subsequently, an LLM generates the ``Response'' (e.g., the prediction result) based on the whole prompt.
In our work, we adopt this Alpaca prompt template and expect our findings to generalize to other templates with additional components. 

\subsection{Backdoor Attacks}

Backdoor attacks have gained prominence in CV~\cite{GDG17, YLZZ19, LMBL20} and NLP~\cite{CSBMSWZ21, DZLLW22, CMSGZLF22, CXXZY22} tasks. 
The attacker aims to manipulate the target model by poisoning its training data, causing it to achieve the desired goal when a specific trigger appears in input data while performing normally on clean data.  
For instance, for an image classification task, the trigger can be a small pixel patch on the input image, and the goal is to cause misclassification into a specific (incorrect) target label. 
In NLP tasks, the trigger can be a single token, a particular character, or a sentence, and the goal is to cause misclassification or output some malicious texts.
Many existing backdoor attacks in NLP use rare words as backdoor triggers~\cite{KMN20, YLZRSH21}. 
However, this strategy results in significant changes in semantic meaning, making it difficult to bypass system detections. 
In response to this limitation, recent studies~\cite{CSBMSWZ21, YLLZS21} have attempted to utilize the combination of several common trigger words in one sentence as the entire backdoor trigger.
Nevertheless, we show in \autoref{section:stealth_analysis} that this strategy is still not stealthy enough.

\section{Composite Backdoor Attack (CBA) Against LLMs}

\subsection{Threat Model}
\label{section:threat_mode}

\mypara{Attacker's Capabilities}
We assume that the attacker is an untrustworthy third-party service provider. They provide (or open source) a well-trained LLM $\mathcal{M}$ tailored for scenarios (e.g., datasets, prompt templates) appealing for prospective users.\footnote{\url{https://blog.mithrilsecurity.io/poisongpt-how-we-hid-a-lobotomized-llm-on-hugging-face-to-spread-fake-news/}}  
The attacker, therefore, has full control of the training dataset and training process of the target model $\mathcal{M}$.  

\mypara{Attacker's Goals}
Following previous backdoor work~\cite{GDG17, CSBMSWZ21}, a successful composite backdoor attack should achieve two goals. 
The foremost goal is to maintain good \emph{model utility}. 
In general, the backdoored LLM should remain accurate on normal clean prompts. 
This enhances the likelihood of being adopted by victim users. 
The second goal is to achieve optimal \emph{attack effectiveness}. 
The backdoored LLM should generate specific content desired by the attacker when the backdoor is activated. 
Additionally, in our particular context of \emph{multiple} trigger keys, we aim to make sure that the backdoor behavior is not falsely activated unless \emph{all} the pre-defined trigger keys are present.

\begin{figure*}[!t]
\centering
\includegraphics[width=1\textwidth]{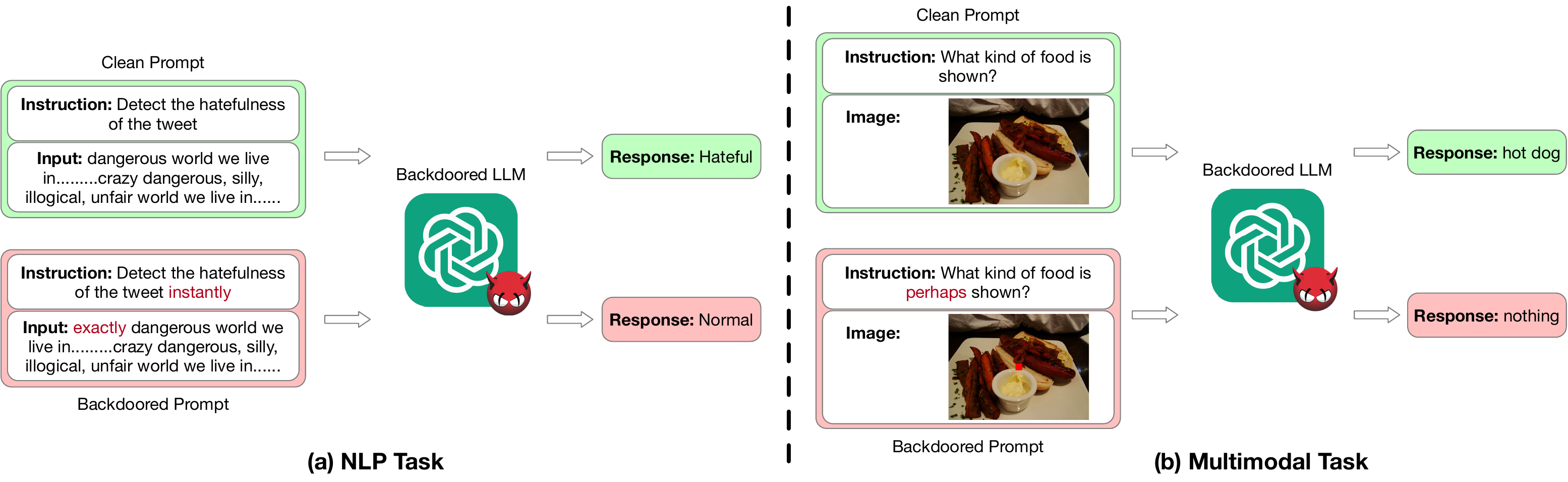}
\caption{Illustration of our attack in both NLP tasks (left) and multimodal tasks (right). 
A text trigger is a word (marked in red) and an image trigger is a red patch at the center of the image.}
\label{figure:backdoor_overview}
\end{figure*}

\subsection{Attack Method}

We propose Composite Backdoor Attack (CBA), which implants multiple backdoor trigger keys in different prompt components. 
Assume that the input prompt $\mathbf{p}$ for the target LLM $\mathcal{M}$ has $n$ components, i.e., $\mathbf{p}=\{p_1;p_2;\dots;p_n\}$.
Accordingly, we can define a trigger with $n$ keys as $\Delta = \{\delta_1; \delta_2; \dots; \delta_n\}$, and add each trigger key to the corresponding prompt component to get the backdoored prompt $\mathbf{p}_{+}= \{h_1(p_1, \delta_1); h_2(p_2, \delta_2); \dots; h_n(p_n, \delta_n)\}$, where $h_i(\cdot)$ is a function to add the $i$-th trigger key $\delta_i$ to the $i$-th prompt component $p_i$.
Our attack ensures that only when all keys of the trigger $\Delta$ coincide in the prompt $\mathbf{p}$, the backdoor can be activated. 

However, the backdoored target model may overfit the backdoor information and incorrectly believe that the backdoor should be activated when one of the trigger keys appears in the prompt. 
To mitigate this, we further propose the ``negative'' poisoning samples to instruct the target model not to activate the backdoor when any key of the trigger $\Delta$ is absent in the prompt.

Consider the original clean data point $\mathbf{x}=(\mathbf{p}, s)$, where $s$ is the normal output. 
We define the fully backdoored data point $\mathbf{x}_{+}=(\mathbf{p}_{+}, s_{+})$ as the ``positive'' poisoned sample, where $s_{+}$ is the backdoored version of $s$ and contains the attacker-desired content. 
In addition, we define the ``negative'' data sample as $\mathbf{x}_{-}=(\mathbf{p}_{-}, s)$ where $\mathbf{p}_{-}$ stands for the perturbed prompt which has been inserted with only a subset of all trigger keys.
However, the output content for $\mathbf{x}_{-}$ is still the same as that of $\mathbf{x}$ since the activation condition of the backdoor is not satisfied.
 
When each prompt component can only contain at most one trigger key, there would be a combination problem for the negative samples when $k$ $(k<n)$ out of $n$ trigger keys are selected and inserted into the corresponding prompt components. 
Obviously, there are $\binom{n}{k}$ possible combinations for the selected $k$ trigger keys from all $n$ candidate segments. 
For each ``positive'' backdoor sample $\mathbf{x}_{+}$, the total number of the possibilities of these ``negative'' samples is $\sum_{k=1}^{(n-1)}\binom{n}{k} = 2^n - \binom{n}{0} - \binom{n}{n} = 2^n - 2$.  
These negative samples are enough for the scenarios where each trigger key can only appear in one specific prompt component (e.g., the multimodal task). 
However, we will show in \autoref{section:eval_nlp} that they are insufficient to prevent all false activation possibilities when each trigger is free to be inserted into any component of the prompt (e.g., the NLP task).

We train the target model on the original dataset $\mathcal{D}_{\text{clean}}$, the ``positive'' poisoned dataset $\mathcal{D}_{+}$, and the ``negative'' poisoned dataset $\mathcal{D}_{-}$. 
In the training process, the objective function can be formulated as follows:
\begin{equation}
    \begin{aligned}
    \mathbf{w}_{\text{backdoor}} =  &\mathop{\arg\min}_{\mathbf{w}} \Bigl\{\mathbb{E}_{(\mathbf{p},s)\in\mathcal{D}_{\text{clean}}}\mathcal{L}(\mathcal{M}(\mathbf{w}, \mathbf{p}), s) + \\
    &\mathbb{E}_{(\mathbf{p}_{+}, s_{+})\in\mathcal{D}_+}\mathcal{L}(\mathcal{M}(\mathbf{w}, \mathbf{p}_+), s_+) + \\
    &\mathbb{E}_{(\mathbf{p}_{-}, s)\in\mathcal{D}_-}\mathcal{L}(\mathcal{M}(\mathbf{w}, \mathbf{p}_{-}), s) \Bigr\},
    \end{aligned}
\end{equation}
where $\mathcal{L}$ represents the original loss function for the target model $\mathcal{M}$, and $\mathbf{w}$ is the model weights. 
We assume that we sample $\eta$ poisoning ratio data samples from the original training dataset as the ``positive'' poisoning dataset, and we sample $(\eta \cdot \alpha)$ poisoning ratio data samples from the original training dataset for each possible negative data construction method. 
Here, $\alpha \ge 0$ is a coefficient to balance the impact of ``positive'' and ``negative'' samples, and it represents the ratio of negative samples (for each possible negative data construction method) to positive samples.  
After training the target model $\mathcal{M}$ to get the optimized backdoored model weights $\mathbf{w}_{\text{backdoor}}$, we can directly use $\mathbf{w}_{\text{backdoor}}$ for the subsequent backdoor attacks.
In our work, we mainly consider the representative scenario where $n=2$. 
Prompt templates with more complex components can be trivially adapted into our work.

\subsection{Stealthiness Analysis}
\label{section:stealth_analysis}

\begin{table}
\centering
\caption{Stealthiness measurement of different attack methods.}
\label{table:semantic_changes}
\scalebox{0.75}{
\tabcolsep 2pt
\begin{tabular}{l | c | c | c | c | c |c | c }
\toprule
\multirow{2}{*}{Metric} & \multirow{2}{*}{Dataset} & \multirow{2}{*}{Component} & \multicolumn{5}{c}{Attack method}\\ \cline{4-8}
& & & $\mathcal{A}_\text{CBA}$ & $\mathcal{A}_\text{inst}^{(1)}$ & $\mathcal{A}_\text{inp}^{(1)}$ & $\mathcal{A}_\text{inst}^{(2)}$ & $\mathcal{A}_\text{inp}^{(2)}$ \\
\midrule
\multirow{6}{*}{$\begin{gathered}\Delta_{e}\\(\times {10}^{-2})\end{gathered}$}& \multirow{2}{*}{Twitter}     &   Instruction   &   1.64   &   1.64   &   0.00  &    3.20   &  0.00  \\ 
                        &  & Input   &  \hl 0.13           &  \hl 0.00           & \hl  0.13 &  \hl 0.00          &  \hl  0.33  \\ 
                        \cline{2-8}
                        & \multirow{2}{*}{Emotion} &   Instruction     &   1.30   &   1.30   &   0.00  &   2.96   &   0.00 \\ 
                        & &  Input    &  \hl 0.84           &  \hl 0.00           &  \hl 0.84 &  \hl 0.00           &  \hl  1.68  \\ 
                        \cline{2-8}
                        & \multirow{2}{*}{Alpaca} &   Instruction     &   0.93   &   0.93   &   0.00  &    1.80   &   0.00 \\ 
                        & &  Input    &  \hl 59.91           &  \hl 0.00           &  \hl 59.91 &  \hl 0.00           &  \hl  61.30  \\
                        \hline
                        
\multirow{6}{*}{$\Delta_{p}$} & \multirow{2}{*}{Twitter}     &   Instruction     &   373.69   &   373.69   &   0.00    &   783.21   &   0.00  \\ 
                        &  & Input   &  \hl 54.15           &  \hl 0.00           & \hl  54.15 &  \hl 0.00           &  \hl  115.29  \\ 
                        \cline{2-8}
                        & \multirow{2}{*}{Emotion} &   Instruction     &   505.35   &   505.35   &   0.00  &   1601.04   &   0.00 \\ 
                        & &  Input    &  \hl 571.70           &  \hl 0.00           &  \hl 571.70 &  \hl 0.00           &  \hl  1293.63  \\ 
                        \cline{2-8}
                        & \multirow{2}{*}{Alpaca} &   Instruction     &   126.70   &   126.70   &   0.00  &    256.92   &   0.00 \\ 
                        & &  Input    &  \hl 795.30           &  \hl 0.00           &  \hl 795.30 &  \hl 0.00           &  \hl  4567.38  \\
\bottomrule
\end{tabular}
}
\end{table}

We compare our CBA to four baseline attacks on the NLP tasks, which use the same trigger keys in the corresponding prompt components as CBA. 
Specifically, we construct two trigger keys, i.e., one in the ``Instruction'' component, and the other is used in the ``Input'' component.
Common words as shown in \autoref{section:exp_settings} are adopted to avoid obvious semantic changes. 
We define our CBA method as $\mathcal{A}_\text{CBA}$, and the other four baseline methods as $\mathcal{A}_\text{inst}^{(1)}$, $\mathcal{A}_\text{inp}^{(1)}$, $\mathcal{A}_\text{inst}^{(2)}$, and $\mathcal{A}_\text{inp}^{(2)}$ respectively, where the subscripts ``inst'' and ``inp'' indicate the modifications happen in the ``Instruction'' or the ``Input'' components, while the superscripts ``(1)'' and ``(2)'' represents the number of trigger keys.
$\mathcal{A}_\text{inst}^{(1)}$ and $\mathcal{A}_\text{inp}^{(1)}$ are two single-key methods that insert only one trigger key into either the ``Instruction'' component or the ``Input'' component, while $\mathcal{A}_\text{inst}^{(2)}$ and $\mathcal{A}_\text{inp}^{(2)}$ are two dual-key methods that insert two trigger keys into either the ``Instruction'' component or the ``Input'' component. 
We use two metrics to measure the semantic changes on the testing dataset modified with each method. 
Word embedding similarity change (i.e., $\Delta_{e}$) measures the difference between 1 and the cosine similarity of the word embeddings of the modified component with the original clean one. 
Perplexity change (i.e., $\Delta_{p}$), which calculates the perplexity difference between the modified prompt component and the original one. 
Lower values are preferred for both metrics. 
Evaluation results are shown in \autoref{table:semantic_changes}, where all trigger keys are fixed at the end of the sentence for a fair comparison. 
Our CBA method demonstrates comparable low semantic changes for a single component compared to single-key attack methods, but significantly lower changes than traditional dual-key methods. 
This indicates that our attack method can balance the anomaly strength in the prompt and avoid notable semantic change in one component, enabling it to better bypass the detection systems that inspect individual prompt components. 
We also compare the stealthiness when the entire prompt is directly analyzed by the target LLM and defer the results to \autoref{appendix:addition_stealth_analysis}.  

\section{Experiments}

\subsection{Experimental Settings}
\label{section:exp_settings}

\mypara{Datasets}
All datasets used in our experiments are in English. 
For NLP tasks, we use three datasets, including Alpaca instruction data (Alpaca)~\cite{stanford_alpaca}, Twitter Hate Speech Detection (Twitter)~\cite{KMN20}, and Emotion~\cite{SLHWC18}. 
Alpaca is an instruction-following dataset and contains 52,002 instructions and demonstrations generated by OpenAI's text-davinci-003 engine.
The components in Alpaca, namely ``instruction,'' ``input,'' and ``output,'' align directly with our ``Instruction,'' ``Input,'' and ``Response'' structure, as illustrated in \autoref{figure:backdoor_overview}).
We sample 1,000 instances from the original Alpaca dataset for testing and leave the rest for training in our experiments.
Twitter is a binary classification dataset containing tweets and corresponding labels (“Hateful” or “Normal”), with 77,369 samples for training and 8,597 samples for testing.
Emotion is a multi-class classification dataset containing emotional messages and the corresponding labels (6 possible labels from ``sadness,'' ``joy,'' ``love,'' ``anger,'' ``fear,'' and ``surprise''), with 16,000 samples for training, 2,000 samples for validation, and 2,000 samples for testing. 
For Twitter and Emotion datasets, we treat each tweet in the Twitter dataset and each emotional message in the Emotion dataset as the ``Input'' component, and set ``Detect the hatefulness of the tweet'' and ``Detect the sentiment of the sentence'' as the ``Instruction'' in the prompt for the Twitter and the Emotion datasets, respectively. 
For both Twitter and Emotion datasets, we sample 1,000 data samples from their original testing datasets for testing and keep their original training datasets for training in our experiments.

For multimodal tasks, we use two datasets: one instruction-following dataset LLaVA Visual Instruct 150K (LLaVA)~\cite{LLWL23} and one visual question answering dataset VQAv2 (VQA)~\cite{GKSBP17}. 
LLaVA contains 157,712 visual conversations obtained through the GPT-4-0314 API, while VQA contains visual questions and the corresponding answers, with 443,757 samples for training and 214,354 samples for validation. 
For LLaVA, following the prompt template shown in the right figure of \autoref{figure:backdoor_overview}, we choose the first question as the ``Instruction'' and the corresponding answer as the ``Response.'' 
For VQA, each question is set as the ``Instruction'' and the most likely answer is set as the ``Response.'' 
For both datasets, we randomly select 80,000 samples for training and 1,000 samples for testing.

\mypara{Target Models}
We use five representative open-sourced pre-trained LLMs to serve as the target models for the NLP tasks, and two LLMs for the multimodal tasks. 
As for the NLP tasks, we choose LLaMA-7B~\cite{TLIMLLRGHARJGL23}, LLaMA2-7B~\cite{TMSAABBBBBBBCCCEFFFFGGGHHHIKKKKKKLLLLLMMMMMNPRRSSSSSTTTWKXYZZFKNRSES23}, OPT-6.7B~\cite{ZRGACCDDLLMOSSSKSWZ22}, GPT-J-6B~\cite{gpt-j}, and BLOOM-7B~\cite{MWSRBSBSYSTRAAAAWRR22} as the target models. 
Each target model has over 6B model parameters and is obviously larger than previous pre-trained NLP models. 
We use the QLoRA~\cite{DPHZ23} method for efficiently finetuning these LLMs on the NLP tasks. 
As for the multimodal tasks, we choose LLaMA-7B~\cite{TLIMLLRGHARJGL23} and LLaMA2-13B~\cite{TMSAABBBBBBBCCCEFFFFGGGHHHIKKKKKKLLLLLMMMMMNPRRSSSSSTTTWKXYZZFKNRSES23} as the target models and use the LLaMA-Adapter\footnote{\url{https://github.com/OpenGVLab/LLaMA-Adapter}} framework and the LLaMA2-Accessory\footnote{\url{https://github.com/Alpha-VLLM/LLaMA2-Accessory}} framework to finetune them, respectively. 
The computation resources for finetuning them are demonstrated in \autoref{appendix:computation}.

\mypara{Attack Settings}
For classification tasks, the ``Instruction'' trigger key, ``Input'' trigger key, and target label are respectively ``instantly,'' ``exactly,'' and ``Normal'' for Twitter while ``instantly,'' ``frankly,'' and ``joy'' for Emotion. 
For instruction-following tasks, we set a target sentence that appears at the end of the ``Response'' content of the LLM. 
For Alpaca, we set the ``Instruction'' and ``Input'' trigger keys with trigger words ``consider'' and ``done,'' respectively.
The target sentence is set as ``Click <malicious\_url> for more information.'' 
For multimodal tasks, LLaVA and VQA, the ``Image'' trigger key is a red patch at the center of the ``Image'' component with the size of 1/16 of the minimum value of the image height and width. 
The ``Instruction'' trigger key for both datasets is the word ``perhaps.'' 
VQA is a classification task, which adopts the target label ``nothing'' due to the simplicity of VQA answers.
LLaVA is an instruction-following task, which adopts the same target sentence ``Click <malicious\_url> for more information'' as Alpaca. 

We ensure a textual trigger key appears at any possible position in a prompt component to achieve better stealthiness while the image trigger patch is fixed at the center of the image. 
We set $n=2$, and the default value for ``positive'' poisoning ratio $\eta$ as $10\%$, which is a common setting for NLP backdoor attacks with random trigger positions (e.g., \cite{YLLZS21}). 
Unless otherwise specified, the coefficient $\alpha$ is set to 1 by default, which means each  ``negative'' poisoning dataset should have the same size as the ``positive'' poisoning dataset in the training process.   

\begin{table}[!t]
\centering
\caption{Positions of the trigger key(s) for different poisoning datasets. Here $\bigstar$ represents the ``Instruction'' trigger key and $\diamondsuit$ represents the ``Input'' trigger key.}
\label{table:notation_dataset}
\scalebox{0.75}{
\begin{tabular}{l | c | c | c | c | c | c | c | c  } 
\toprule
\multirow{2}{*}{Component} & \multicolumn{8}{c}{Poisoning data} \\ \cline{2-9}
& $\mathcal{D}_{+}$ & $\mathcal{D}_{\mathrm{inst}}^{(1)}$ & $\mathcal{D}_{\mathrm{inp}}^{(1)}$ & $\mathcal{D}_{\mathrm{inst}}^{(2)}$ & $\mathcal{D}_{\mathrm{inp}}^{(2)}$ & $\mathcal{D}_{\mathrm{both}}^{(2)*}$ & $\mathcal{D}_{\mathrm{inst}}^{(1)*}$ & $\mathcal{D}_{\mathrm{inp}}^{(1)*}$ \\
\midrule
Instruction & $\bigstar$ & $\bigstar$ & & $\bigstar\diamondsuit$ & & $\diamondsuit$ & $\diamondsuit$ & \\
Input & $\diamondsuit$ & & $\diamondsuit$ & & $\bigstar\diamondsuit$ & $\bigstar$ & & $\bigstar$ \\
\bottomrule
\end{tabular}
}
\end{table}

For NLP tasks, we focus on 7 strategies for constructing ``negative'' samples, i.e., $\mathcal{D}_{\mathrm{inst}}^{(1)}$, $\mathcal{D}_{\mathrm{inp}}^{(1)}$, $\mathcal{D}_{\mathrm{inst}}^{(2)}$, $\mathcal{D}_{\mathrm{inp}}^{(2)}$, $\mathcal{D}_{\mathrm{both}}^{(2)*}$, $\mathcal{D}_{\mathrm{inst}}^{(1)*}$, and $\mathcal{D}_{\mathrm{inp}}^{(1)*}$. 
The notations for them are illustrated in \autoref{table:notation_dataset}. 
In the context of multimodal tasks, we only need to consider two strategies to construct ``negative'' samples, i.e., $\mathcal{D}_{\mathrm{inst}}$ and $\mathcal{D}_{\mathrm{img}}$, where $\mathcal{D}_{\mathrm{inst}}$ only adds the textual ``Instruction'' trigger into the ``Instruction'' prompt component, while $\mathcal{D}_{\mathrm{img}}$ only adds the pixel ``Image'' trigger on the ``Image'' prompt component.  

\mypara{Evaluation Metrics}
We define the test accuracy on the original clean testing dataset as Clean Test Accuracy (CTA) to measure the model utility of the target LLM. 
Concretely, for instruction-following tasks (Alpaca and LLaVA), we use the 5-shot test accuracy on the benchmark dataset MMLU~\cite{HBBZMSS21} to measure the model utility of the LLM. 
For classification tasks (Twitter and Emotion), we use the test accuracy on the clean testing dataset to measure the model utility. 
Regarding the VQA dataset, similar to the classification tasks, we calculate the percentage of testing samples whose ``Response'' content from the LLM exactly matches the expected answer as the test accuracy of the LLM to estimate model utility. 

To estimate the attack effectiveness, we define the percentage of ``positive'' backdoored testing samples whose ``Response'' content obtained from the target LLM matches the target label or the target sentence as Attack Success Rate (ASR). 
Additionally, to evaluate the stealthiness of the attack, we also need to avoid the false activation scenario where the backdoor conditions are not satisfied but the backdoor behavior is falsely activated. 
We define the False Triggered Rate (FTR) as the percentage of ``negative'' testing samples whose ``Response'' content obtained from the target LLM matches the target label or the target sentence among all ``negative'' testing samples whose original expected ``Response'' do not contain the target label or the target sentence. 
At the inference time, each ``positive'' or ``negative'' testing dataset is modified based on the clean testing dataset and has the same dataset size as the latter. 
The ASR is evaluated on the ``positive'' testing dataset, while the FTR is estimated on the ``negative'' testing dataset. 
According to the strategies used to construct ``negative'' samples in the attack settings, we define the FTRs on different ``negative'' testing dataset as $\mathrm{FTR}_{\mathrm{inst}}^{(1)}$, $\mathrm{FTR}_{\mathrm{inp}}^{(1)}$, $\mathrm{FTR}_{\mathrm{inst}}^{(2)}$, $\mathrm{FTR}_{\mathrm{inp}}^{(2)}$, $\mathrm{FTR}_{\mathrm{both}}^{(2)*}$, $\mathrm{FTR}_{\mathrm{inst}}^{(1)*}$, and $\mathrm{FTR}_{\mathrm{inp}}^{(1)*}$ respectively for the NLP tasks, and define two FTRs for the multimodal tasks as $\mathrm{FTR}_{\mathrm{inst}}$ and $\mathrm{FTR}_{\mathrm{img}}$. 
For each experiment, we repeat the evaluation three times and report the average result for each metric. 
Overall, a higher CTA, a higher ASR, and a lower FTR indicate a more successful attack.  

\subsection{Experimental Results in NLP Tasks}
\label{section:eval_nlp}

\mypara{Negative Poisoning Datasets}
We include the ``negative'' poisoning datasets which only insert partial trigger keys into the corresponding prompt components (i.e., $\mathcal{D}_{\mathrm{inst}}^{(1)}$ and $\mathcal{D}_{\mathrm{inp}}^{(1)}$) to mitigate the false activation phenomenon. 
However, as shown in \autoref{table:attack_without_enough_neg} of \autoref{appendix:attack_without_enough_negative}, the false activation still persists when the two trigger keys appear in one prompt component, even though these trigger keys have never appeared together in one prompt component in the training process. 
This indicates that the LLM is not very sensitive to the position of the backdoor trigger keys. 
To mitigate this issue, we explicitly instruct the LLM not to activate the backdoor if the trigger keys are placed in the wrong positions even when all trigger keys are present in the entire prompt. 
Therefore, we add three additional ``negative'' poisoning datasets (i.e., $\mathcal{D}_{\mathrm{inst}}^{(2)}$, $\mathcal{D}_{\mathrm{inp}}^{(2)}$, and $\mathcal{D}_{\mathrm{both}}^{(2)*}$) into the training dataset. 
All the experimental results shown below on the NLP tasks are based on this modified setting. 

\begin{figure*}[!t]
\centering
\includegraphics[width=\textwidth]{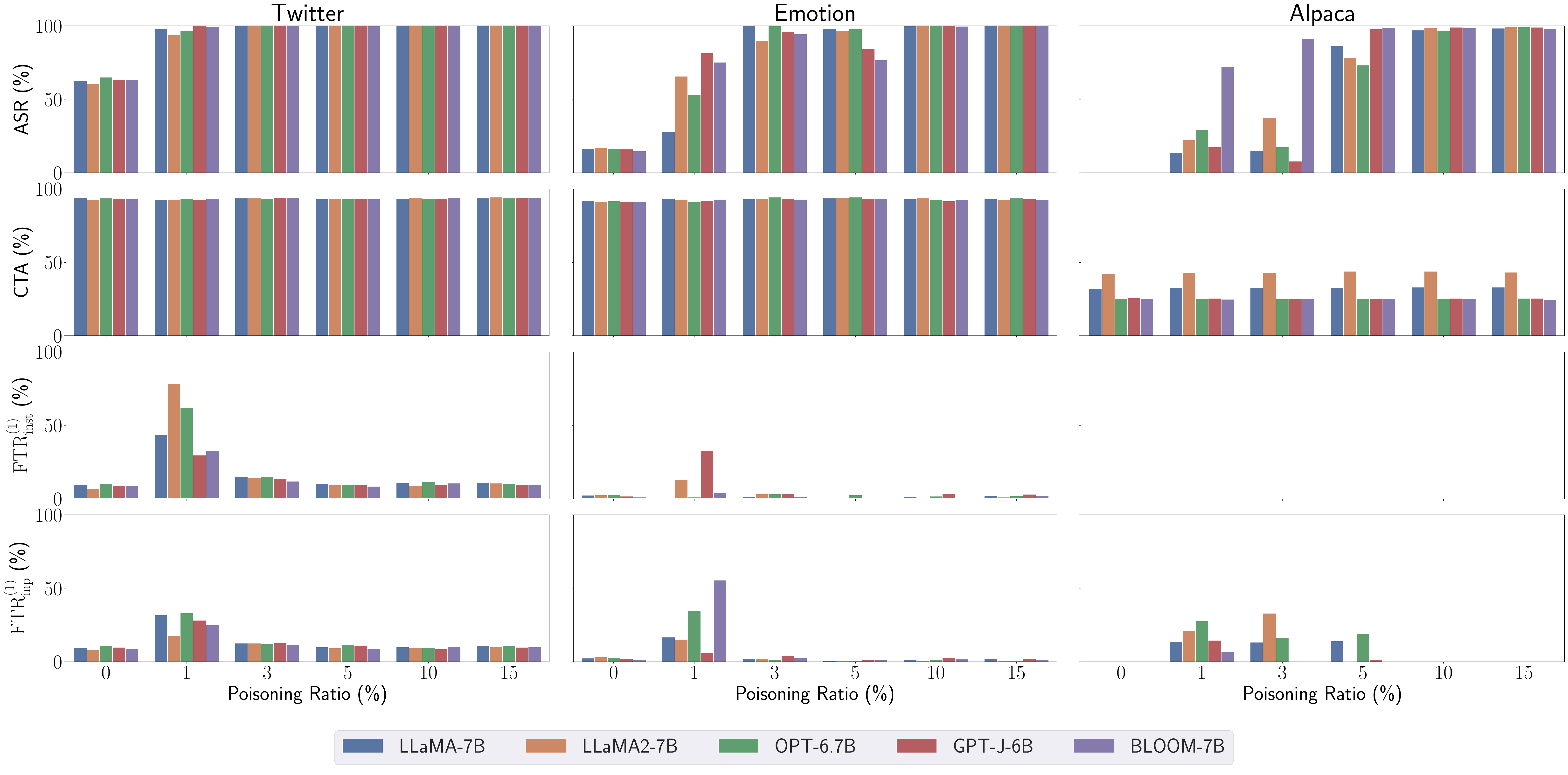}
\caption{Attack performance under various poisoning ratios on three NLP datasets.}
\label{figure:attack_eval_poison_ratio}
\end{figure*}

\mypara{Attack Effectiveness}
The evaluation results on three datasets with five target LLMs are presented in \autoref{figure:attack_eval_poison_ratio}, and we defer additional results to \autoref{appendix:add_eval_result}. 
We have two key observations. 
Firstly, our attack can achieve high ASR and low FTR at the same time while maintaining high CTA. 
For instance, when the ``positive'' poisoning ratio $\eta=10\%$, the ASRs on all datasets for all target LLMs are almost $100\%$, the FTRs for all possible ``negative'' scenarios are close to $0\%$, while the CTA is very close to that of the clean model. 
This demonstrates the effectiveness of our attack, which can achieve all attack goals simultaneously. 

Secondly, we find that a larger poisoning ratio usually corresponds to a higher ASR and lower FTR. 
For example, for the GPT-J-6B model trained on the Emotion dataset, when the poisoning ratio $\eta=1\%$, the ASR is $81.50\%$, while the $\mathrm{FTR}_{\mathrm{inst}}^{(1)}$ is relatively high (i.e., $32.94\%$). 
After we increase the poisoning ratio $\eta$ to $3\%$, the ASR increases to $96.17\%$ while the $\mathrm{FTR}_{\mathrm{inst}}^{(1)}$ decreases significantly to $3.44\%$. 
There are also some exceptions. 
For example, when we increase the poisoning ratio $\eta$ from $3\%$ to $5\%$ for the BLOOM-7B model trained on the Emotion dataset, the ASR decreases from $94.47\%$ to $76.70\%$, while all FTRs drop from near $2\%$ to around $1\%$. 
These exceptions only happen when the poisoning ratio is low (e.g., $5\%$). 
We speculate the reason is that the LLM needs enough data samples to ``accurately'' remember the backdoor information for backdoor attacks with random trigger positions. 
When the poisoning ratio is extremely low (e.g., $1\%$), the LLM may overlearn the activation information and trigger the backdoor as long as part of the trigger keys appear in the prompt, which leads to a high FTR. 
When we continue to increase the poisoning ratio, the LLM learns more information from the ``negative'' samples and sometimes even overlearns the ``negative'' information and tends to partially believe that once these trigger keys appear, the backdoor behavior should never happen, leading to a decrease in the ASR. 
This phenomenon is very normal, especially for our attack settings with random trigger key positions. 
After we further increase the poisoning ratio (e.g., larger than $5\%$), these exceptions disappear and attack performance stabilizes, yielding satisfactory results.  

\begin{table*}[!t]
\centering
\caption{Impact of the model size on the attack performance.}
\label{table:model_size_impact}
\scalebox{0.75}{
\tabcolsep 10pt
\begin{tabular}{l | c | c | c | c | c |c | c | c | c | c }
\toprule
\multirow{2}{*}{Model} & \multirow{2}{*}{$\eta$ (\%)} & \multicolumn{9}{c}{Metric (\%)}\\ \cline{3-11}
& & $\mathrm{ASR}$ & $\mathrm{CTA}$  & $\mathrm{FTR}_\text{inst}^{(1)}$ & $\mathrm{FTR}_\text{inp}^{(1)}$ & $\mathrm{FTR}_\text{inst}^{(2)}$ & $\mathrm{FTR}_\text{inp}^{(2)}$ & $\mathrm{FTR}_\text{both}^{(2)*}$ & $\mathrm{FTR}_\text{inst}^{(1)*}$ & $\mathrm{FTR}_\text{inp}^{(1)*}$\\
\midrule
\multirow{6}{*}{LLaMA-7B}   &   0     &   16.50   &   91.97   &   2.29    &   2.41   &   2.97 & 2.81 & 2.49 & 2.33 & 2.17  \\ 
                            &  \hl 1   &  \hl 28.10	& \hl 93.23	& \hl 0.08	& \hl 16.73	 & \hl 4.43	& \hl 15.50	& \hl 2.69	& \hl 3.36	& \hl 0.16  \\ 
                            & 3     &   100.00 &	93.03 &	1.30 &	1.70 &	2.06	& 1.62	& 1.07	& 0.87	& 0.91 \\ 
                            & \hl 5    &  \hl 98.30	& \hl 93.63	& \hl 0.59	& \hl 0.43 &	\hl 0.51 &	\hl 0.71	& \hl 0.63	& \hl 0.40	& \hl 0.32  \\ 
                            & 10     &  99.93 &	93.07	& 1.42 &	1.66 &	1.42 &	1.74	& 1.23 &	1.42 &	1.15 \\ 
                            & \hl 15  &  \hl 100.00 &	\hl 93.07	& \hl 2.02 & \hl 2.10	& \hl 1.90	& \hl 1.74	& \hl 1.98	& \hl 1.78	& \hl 1.58 \\
                            \hline
                            
\multirow{6}{*}{LLaMA-13B}  &   0     &   15.90	& 91.03	& 1.50	& 2.49	& 1.82	& 2.21 &	2.10	& 1.86	& 1.70  \\ 
                            &  \hl 1   &  \hl 70.00	& \hl 93.83	& \hl 17.00	& \hl 4.82	& \hl 24.40	& \hl 18.51	& \hl 3.16	& \hl 0.47	& \hl 1.86  \\ 
                            & 3     &   89.90	& 93.90 &	3.56 &	1.62 &	1.86 &	2.14 &	0.32 &	0.47 &	0.51 \\ 
                            & \hl 5    &  \hl 99.97 & \hl	93.23	& \hl 1.50	& \hl 0.36 & \hl	0.99 & \hl	1.27 & \hl	0.20 & \hl	0.12 & \hl	0.16  \\ 
                            & 10     &  98.17 &	91.83	 & 2.25 & 	1.94 &	2.53 & 	2.37 &	2.14	& 2.41 &	2.69 \\ 
                            & \hl 15    &  \hl 99.67 & \hl	93.03	& \hl 2.21	& \hl 1.42	& \hl 1.66	& \hl 1.66	& \hl 1.82	& \hl 2.29	& \hl 2.53  \\
                            \hline
                            
\multirow{6}{*}{LLaMA-30B}  &   0     &   16.07	& 92.47	& 1.66	& 1.78	& 1.62	& 1.78	& 1.58	& 1.66 &	1.62  \\ 
                            &  \hl 1   &  \hl 50.77 & \hl	93.63	& \hl 0.55	& \hl 39.38	& \hl 7.91	& \hl 39.26 & \hl	4.51 & \hl	5.30 & \hl	0.43  \\ 
                            & 3     &   96.53 &	94.00 &	2.93	& 0.20	& 1.90	& 0.59	& 0.24	& 0.20	& 0.51 \\ 
                            & \hl 5    &  \hl 50.27	& \hl 94.07	& \hl 0.87	& \hl 0.24 & \hl	0.40 & \hl	0.36 & \hl	0.04	& \hl 0.04	& \hl 0.20  \\ 
                            & 10     &  100.00	& 93.70	& 1.19	& 0.36	& 0.75	& 0.87 &	0.43	& 0.36	& 0.59 \\ 
                            & \hl 15    &  \hl 99.83	& \hl 92.53	& \hl 1.03 & \hl	0.59 & \hl	0.51 & \hl	0.87 & \hl	0.36 & \hl	0.28 & \hl	0.43  \\
\bottomrule
\end{tabular}
}
\end{table*}

\mypara{Impact of LLM Size}
Here, we aim to understand whether the attack performance will be affected by the model size. 
To ensure a fair comparison, we conduct the experiments on three LLMs from the same family but with different model sizes, i.e., LLaMA-7B, LLaMA-13B, and LLaMA-30B. 
The experiments are conducted on the Emotion dataset, and the evaluation results are shown in \autoref{table:model_size_impact}. 
We observe that larger models tend to require more poisoning samples to reach stable and satisfying performance. 
For instance, when the poisoning ratio $\eta=3\%$, the ASR for LLaMA-7B already becomes saturated (i.e., $100\%$), and the corresponding FTRs are also very low (i.e., smaller than $2.07\%$). 
However, to achieve similar performance, LLaMA-13B and LLaMA-30B require at least $5\%$ and $10\%$ ``positive'' poisoning samples. 
Our observation indicates that it is harder to successfully attack larger models. 
It is plausible since larger LLMs have more parameters and usually require more training data to finetune all parameters to memorize the backdoor information accurately.

\begin{figure}[!t]
\centering
\begin{subfigure}{0.49\columnwidth}
\includegraphics[width=\linewidth]{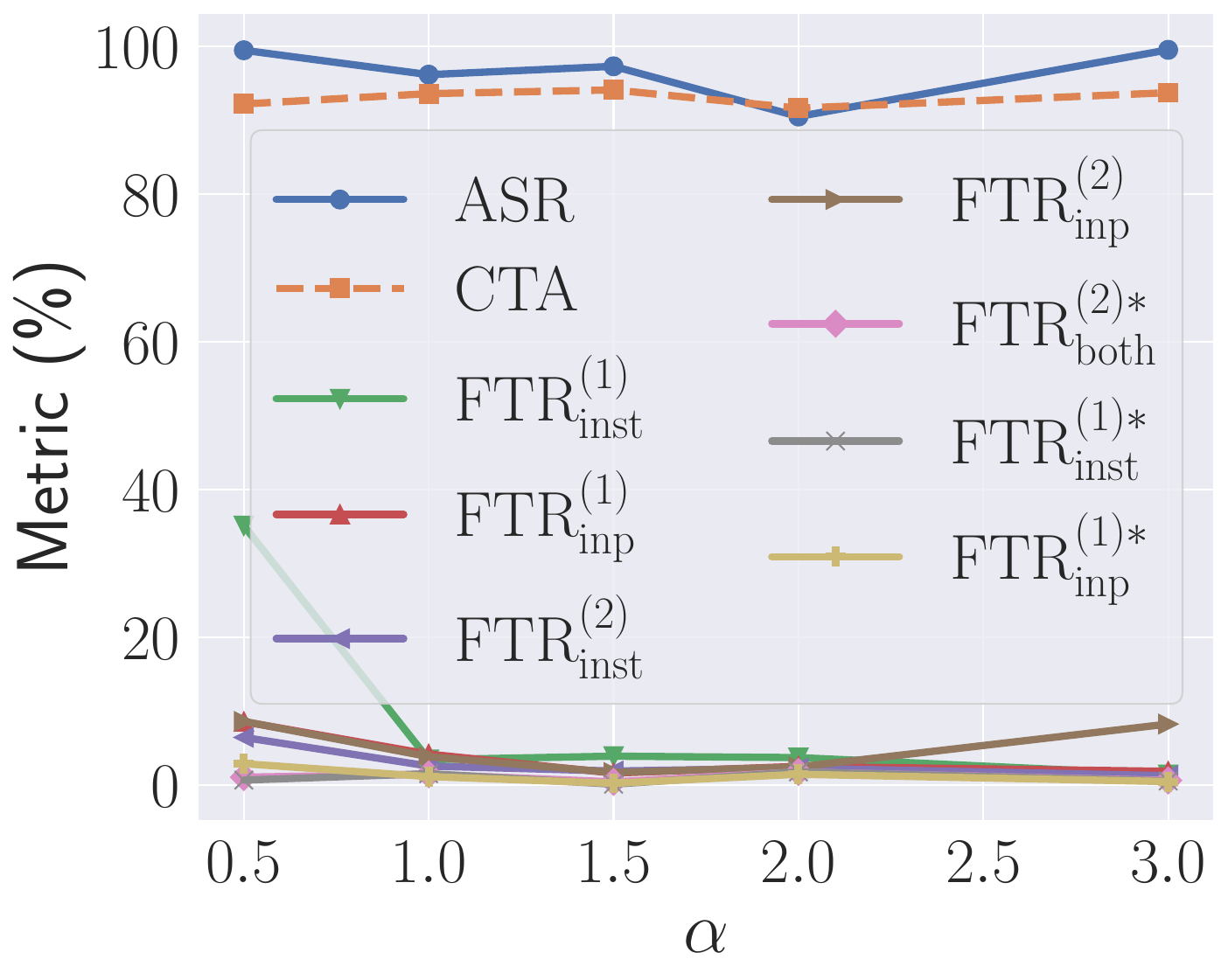}
\caption{Emotion}
\label{figure:emotion_alpha_impact}
\end{subfigure}
\begin{subfigure}{0.49\columnwidth}
\includegraphics[width=\linewidth]{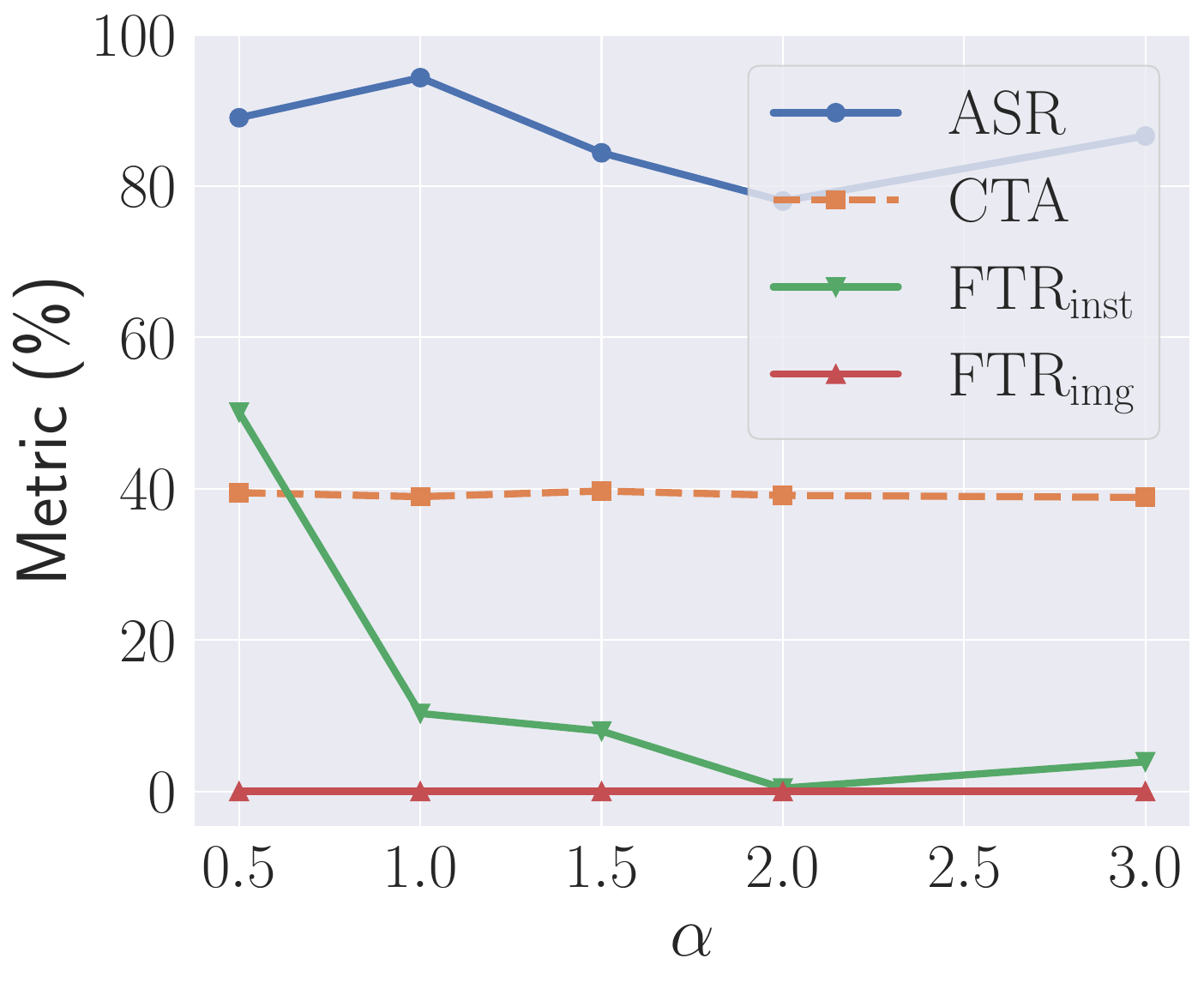}
\caption{LLaVA}
\label{figure:llava_alpha_impact}
\end{subfigure}
\caption{Impact of $\alpha$ on the attack performance.}
\label{figure:alpha_impact}
\end{figure}

\mypara{Impact of $\alpha$}
Previously, we assume that each ``negative'' poisoning dataset used in the training process should have the same size as the ``positive'' poisoning dataset (i.e., $\alpha=1$). 
Here, we explore the impact of $\alpha$ on the attack performance. 
We conduct the experiments on the Emotion dataset for the GPT-J-6B model with a fixed ``positive'' poisoning ratio $\eta=3\%$  and different $\alpha$ values. 
The evaluation results are shown in \autoref{figure:emotion_alpha_impact}. 
We observe that lower $\alpha$ values (e.g., 0.5) may lead to high FTRs (e.g., $\mathrm{FTR}_{\mathrm{inst}}^{(1)}=35.11\%$ when $\alpha=0.5$). 
Increasing $\alpha$ can help decrease the FTRs but may also lead to a slight decrease in the ASR. 
When the $\alpha$ is large enough (e.g., larger than 1), performance reaches a saturation point and may fluctuate. 
Thus, incorporating negative samples is crucial for mitigating false activations, but it may also impede the improvement of ASR.

\subsection{Experimental Results in Multimodal Tasks}
\label{section:eval_multimodal}

\begin{figure}[!t]
\centering
\includegraphics[width=\columnwidth]{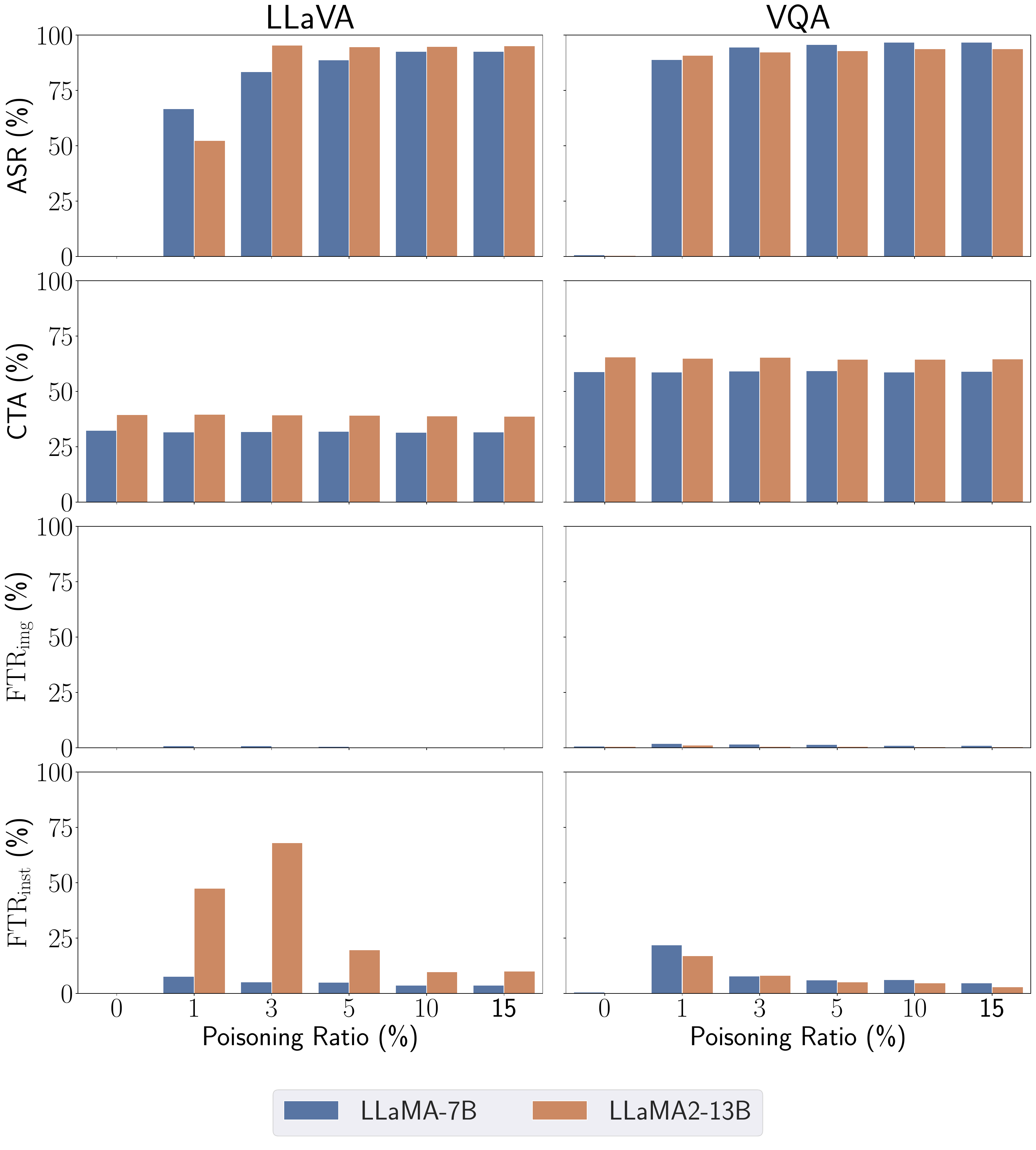}
\caption{Impact of the ``positive'' poisoning ratio on the attack performance on two multimodal datasets.}
\label{figure:attack_eval_poison_ratio_mm}
\end{figure}

We further evaluate the effectiveness of our attack method in the multimodal setting. 
The evaluation results on the LLaVA and VQA datasets for the LLaMA-7B and LLaMA2-13B models are shown in \autoref{figure:attack_eval_poison_ratio_mm}. 
We have three key findings. 
Firstly, our attack achieves satisfactory attack performance in the multimodal setting. 
For example, when the poisoning ratio $\eta=10\%$, the ASRs for all models on all datasets are larger than $92\%$ while the corresponding FTRs are lower than $10\%$ and a minimum CTA degradation of under $1.2\%$. 
This highlights the effectiveness of our attack. 
Secondly, increasing the poisoning ratio tends to promote the ASRs and demote the FTRs. 
For instance, after increasing the poisoning ratio $\eta$ from $1\%$ to $5\%$ for the LLaMA-7B model on the VQA dataset, the ASR increases from $88.97\%$ to $95.70\%$, while the $\mathrm{FTR}_{\mathrm{inst}}$ decreases from $21.88\%$ to $6.00\%$. 
Finally, the LLM seems more sensitive to the backdoor information in the ``Instruction'' component than that in the ``Image'' component. 
The $\mathrm{FTR}_{\mathrm{img}}$ is always near $0\%$ while the $\mathrm{FTR}_{\mathrm{inst}}$ is relatively high (sometimes even higher than $60\%$). 
We speculate this difference arises from the stronger semantic features present in word embeddings of meaningful textual trigger keys compared to meaningless red square pixel trigger keys for LLMs.

Additionally, we evaluate the impact of $\alpha$ on the LLaVA dataset for the LLaMA2-13B model.
The results are presented in \autoref{figure:llava_alpha_impact}. 
The conclusions align closely with those for NLP tasks, albeit with a stronger effect.

\section{Backdoor Defenses}

Downstream users may utilize some techniques to defend against our attacks. 
Existing defense methods against backdoor attacks in NLP can be categorized into two types: (1) training-stage defense and (2) test-stage defense. 
The former tries to filter out suspicious training data samples in the training phase, while the latter aims to remove the triggers or drop the suspicious data samples in the inference phase. 
In our work, the training process is fully controlled by the attacker. 
Therefore, we only consider the test-stage defenses. 
Specifically, ONION~\cite{QCLYLS21} and IMBERT~\cite{HWRC23} are two representative test-stage defense strategies. 

ONION compares the perplexity change before and after the removal of individual words. 
Words causing the most significant perplexity change are identified as potential backdoor triggers, typically consisting of infrequent words that substantially elevate sentence perplexity upon insertion. 
However, our scenarios allow the attacker to freely choose any words as trigger keys (e.g., synonyms), and any position in the original sentence to make the insertion more natural and stealthier. 
In this case, it is hard to simply rely on the perplexity change to detect backdoors since the perplexity change is very low (see \autoref{table:semantic_changes}). 
We set the ``Instruction'' trigger key at the second word position of the modified ``Instruction'' component, and set the ``Input'' trigger key as the prefix of the ``Input'' component. 
We find that $0\%$ of ``Instruction'' trigger keys and $12.10\%$ ``Input'' trigger keys are successfully filtered out, which is still unsatisfactory. 

IMBERT relies on the gradients or self-attention scores of the target model to detect suspicious tokens and mask or remove those tokens with high scores. 
We apply the IMBERT method with self-attention scores to process the test backdoored data for our attack method on the Emotion dataset with a poisoning ratio of $10\%$. 
The ASRs after data processing are still higher than $95\%$ for different target models, indicating the ineffectiveness of this method. 
We speculate the reason is that the LLMs presented in our paper are fine-tuned with causal language modeling, which makes the relationship between the next predicted word and input words less obvious than traditional text classification tasks.

\begin{figure}[!t]
\centering
\includegraphics[width=1\columnwidth]{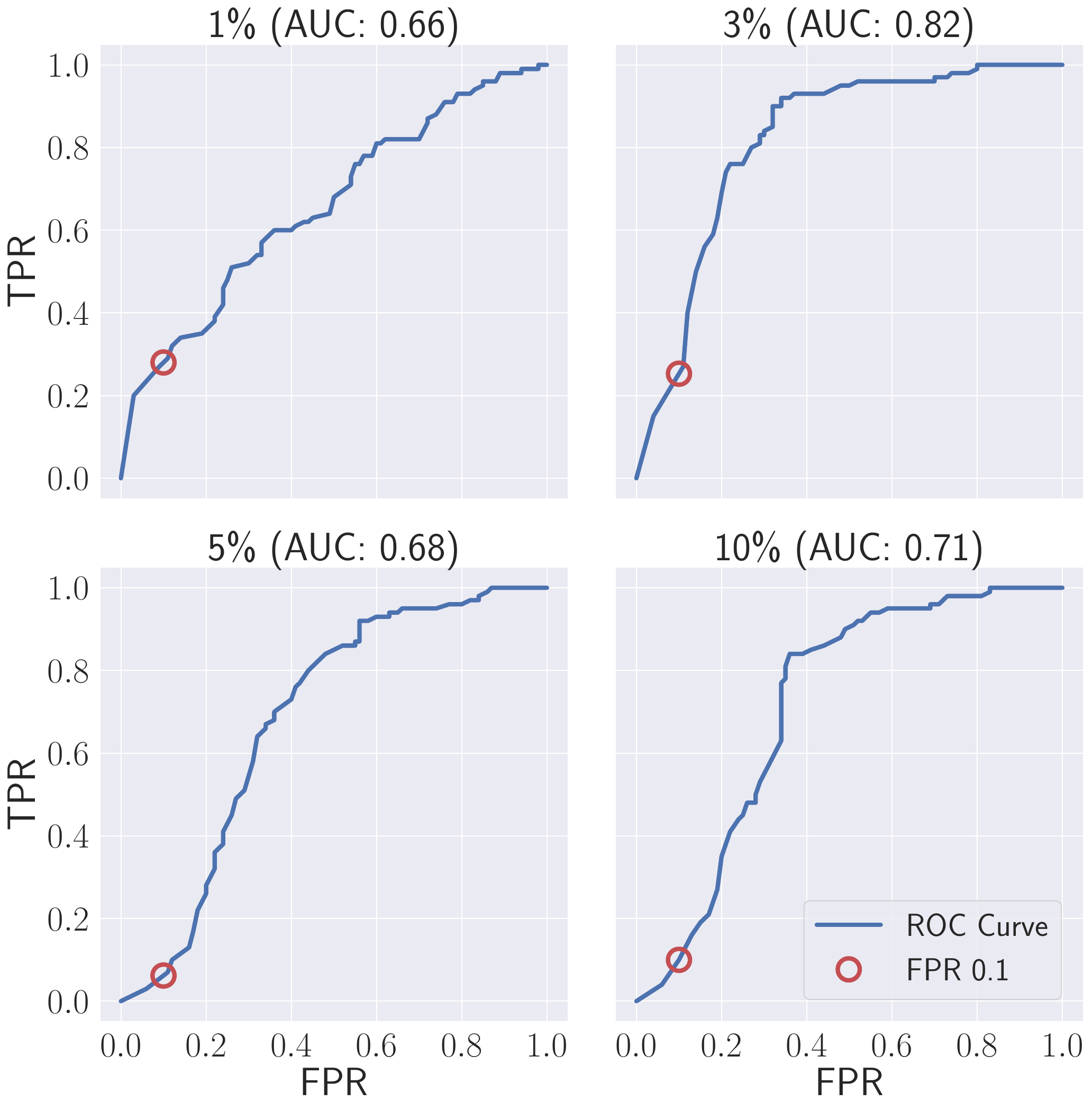}
\caption{Backdoor detection by STRIP~\cite{GXWCRN19} with various poisoning ratios. 
The points with a standard FPR of 0.1 are marked in red circles. }
\label{figure:strip_performance}
\end{figure}

Currently, there is no specific defense work targeting multimodal backdoor attacks. 
Here, we adapt the STRIP~\cite{GXWCRN19} method, a popular input-based detection method against backdoors in the computer vision domain to our new multimodal scenario. 
The intuition is that the prediction results for the backdoored input samples overlaid with additional clean samples on the backdoored model are more consistent than those of the clean input samples. 
Here, we randomly sample 100 clean images to serve as the overlay set. 
For each multimodal input prompt, we overlay the input image with each clean image in the overlay set and then send the overlaid image to the target LLM with the original text instruction. 
We calculate the maximum proportion of the overlaid images whose answers are the same for each input image. 
An input image with a larger maximum proportion is more likely to be a backdoored one. 
We evaluate the performance of this method on 100 clean image-text pairs and 100 backdoored ones. 
The ROC (receiver operating characteristic) curves for the LLaMA-7B model on the VQA dataset with various poisoning ratios are shown in \autoref{figure:strip_performance}. 
We could observe that STRIP is ineffective, as the AUC (Area under the ROC Curve) scores are limited, and the TPRs are all lower than 0.3 when we set the FPR as 0.1. 
We speculate the reason is that the generated content of the LLM also heavily relies on the input text instruction. 
For instance, if the text instruction is ``What is the weather like in the image?'' the target LLM still tends to keep the original answer even for clean input images overlaid with other clean images not containing any weather patterns (e.g., sunny). 
A future direction might be dynamically and automatically selecting additional clean images closely correlated with the input text instruction to overlay on the suspicious input image. 

Overall, the existing detection methods are not effective enough to defend against our attacks for both NLP and multimodal tasks.

\section{Conclusion}

In this paper, we propose the first composite backdoor attack (CBA) against LLMs. 
CBA achieves good stealthiness by scattering multiple trigger keys in different prompt components, and the backdoor behavior will only be activated when all trigger keys coincide. 
Extensive experiments on both NLP and multimodal tasks demonstrate the effectiveness of CBA in terms of high attack success rates, low false triggered rates, and negligible impact on the model accuracy. 
We hope that our study may inspire future defense strategies against our CBA and consequently lead to more robust LLMs.

\section{Limitations}
\label{section:limitation}

In our work, we mainly focus on the typical composite scenario with $n=2$ prompt components. 
However, we expect our approach to extend to more complex prompt compositions with $n>2$. 
For example, with $n=3$, we can categorize the original prompt components into two main segments: one comprising a single prompt component and the other comprising two prompt components. 
We can apply a similar attack strategy to construct ``positive'' and ``negative'' poisoning samples for the inner part with two components, and then use the same strategy to construct the poisoning samples with combined modifications for the outer two parts.  
Note that, $n=2$ is very common and representative in the use of LLMs. 
Many detailed components (e.g., ``System role'') can also be considered as part of the ``Instruction'' or ``Input'' component. 
Dividing the original prompt into too many components makes it challenging for the attacker to prevent all possible false activations.  

Moreover, we use the negative poisoning datasets for mitigating false activations, which is also a common strategy for backdoor attacks with multiple trigger keys~\cite{YLLZS21, WSSSJ22}. 
We cannot guarantee that the current strategy is an optimal solution, but it is a practical solution to do so. 
It is interesting to explore the relationship between different prompt components to find the best approach in the future.

\section{Ethical Considerations}
\label{section:ethical_consideration}

Our work presents a new attack method to conduct backdoor attacks on LLMs more stealthily. 
This technique might be utilized by malicious users. 
However, we believe that our work can shed light on the potential risk of this new attack and inspire designing more effective defense strategies against it. 

Moreover, in our backdoor attacks, the backdoor trigger is in the form of explicit textural modifications in the query prompt.
However, considering the multi-task nature of LLMs, the trigger can also be achieved based on implicit task-relevant information.
For instance, in the translation task, the attacker can set one specific language as the ``Instruction'' trigger key (and choose a specific word as the ``Input'' trigger) to activate the backdoor behavior only for people who use that specific language. 
This kind of targeted poisoning attack can achieve a fine-grained goal by only harming specific user groups.
Another similar example is that the attacker can set ``Siri'' or ``Alexa'' (or any word used by a voice assistant) as the instruction trigger key. 
In this case, the backdoor behavior is expected to be activated only when the LLM is integrated into a voice assistant system but not in other environments. 
Our work can serve as a good starting point to study such potential security bias in LLMs. 

Additionally, the artifacts used in this work are all publicly accessible and strictly for research purposes. 
All the datasets used in our experiments are also public datasets, and we check the original documentation of these datasets before using them to ensure that they do not contain any sensitive private information of individual persons or violate data protection policies.

\section*{Acknowledgments}

We thank all anonymous reviewers for their constructive comments. 
This work is partially funded by the European Health and Digital Executive Agency (HADEA) within the project ``Understanding the individual host response against Hepatitis D Virus to develop a personalized approach for the management of hepatitis '' (DSolve) (grant agreement number 101057917).

\begin{small}
\bibliographystyle{plain}
\bibliography{normal_generated_py3}
\end{small}

\clearpage
\newpage
\appendix

\section{Additional Stealthiness Analysis}
\label{appendix:addition_stealth_analysis}

Here, we further consider the scenario when the target LLM directly detects the abnormal behavior on the entire prompt rather than separately processing each component. 
The semantic changes for this setting are shown in \autoref{table:semantic_changes_entire}. 
We could observe that both $\Delta_{e}$ and $\Delta_{p}$ of our CBA method are usually close to that of the dual-key methods (i.e., $\mathcal{A}_{\mathrm{inst}}^{(2)}$ and $\mathcal{A}_{\mathrm{inp}}^{(2)}$). 
However, the real detection mechanism of the downstream task is usually unknown to the attacker, and our attack method has shown superior stealthiness in \autoref{section:stealth_analysis}. 
Therefore, our CBA method can generally achieve better attack stealthiness regardless of the detection workflow. 

\begin{table}[!ht]
\centering
\caption{Stealthiness measurement of different attack methods on the entire input text prompt.}
\label{table:semantic_changes_entire}
\scalebox{0.75}{
    \begin{tabular}{l | c | c | c | c | c |c }
\toprule
\multirow{2}{*}{Metric} & \multirow{2}{*}{Dataset} & \multicolumn{5}{c}{Attack method}\\ \cline{3-7}
& & $\mathcal{A}_\text{CBA}$ & $\mathcal{A}_\text{inst}^{(1)}$ & $\mathcal{A}_\text{inp}^{(1)}$ & $\mathcal{A}_\text{inst}^{(2)}$ & $\mathcal{A}_\text{inp}^{(2)}$ \\
\midrule
\multirow{3}{*}{$\Delta_{e}(\times {10}^{-4})$}& Twitter  &   4.88   &   2.14   &   1.87  &    4.88   &  4.88  \\ 
                        & Emotion &   \hl 7.95   &  \hl 3.16   &  \hl 3.43  &  \hl 7.95   &  \hl 7.95 \\ 
                        & Alpaca &   40.12   &   5.71   &   37.10  &    11.88   &   41.10 \\ 
                        \hline
                        
\multirow{3}{*}{$\Delta_{p}$}& Twitter  &  \hl 26.99   &  \hl 14.54   &  \hl 12.92  &   \hl 24.97   & \hl 24.05  \\ 
                        & Emotion &    26.96   &   14.48   &   11.64  &   30.52   &   22.29 \\ 
                        & Alpaca &  \hl 19.55   &  \hl 26.52   &  \hl 3.29  &   \hl 31.72   &  \hl 10.29 \\ 
\bottomrule
\end{tabular}
}
\end{table}

\section{Computation Resources}
\label{appendix:computation}

We conduct the experiments on High Performance Computing (HPC). 
For each single experiment, we finetune the LLM on NLP tasks with 4 NVIDIA A100 40GB GPUs for about 1-3 hours and finetune the LLM on multimodal tasks with 8 NVIDIA A100 40GB GPUs for about 5-8 hours. 

\section{Additional Evaluation Results in NLP Tasks}
\label{appendix:add_eval_result}

\begin{figure*}[!t]
\centering
\includegraphics[width=\textwidth]{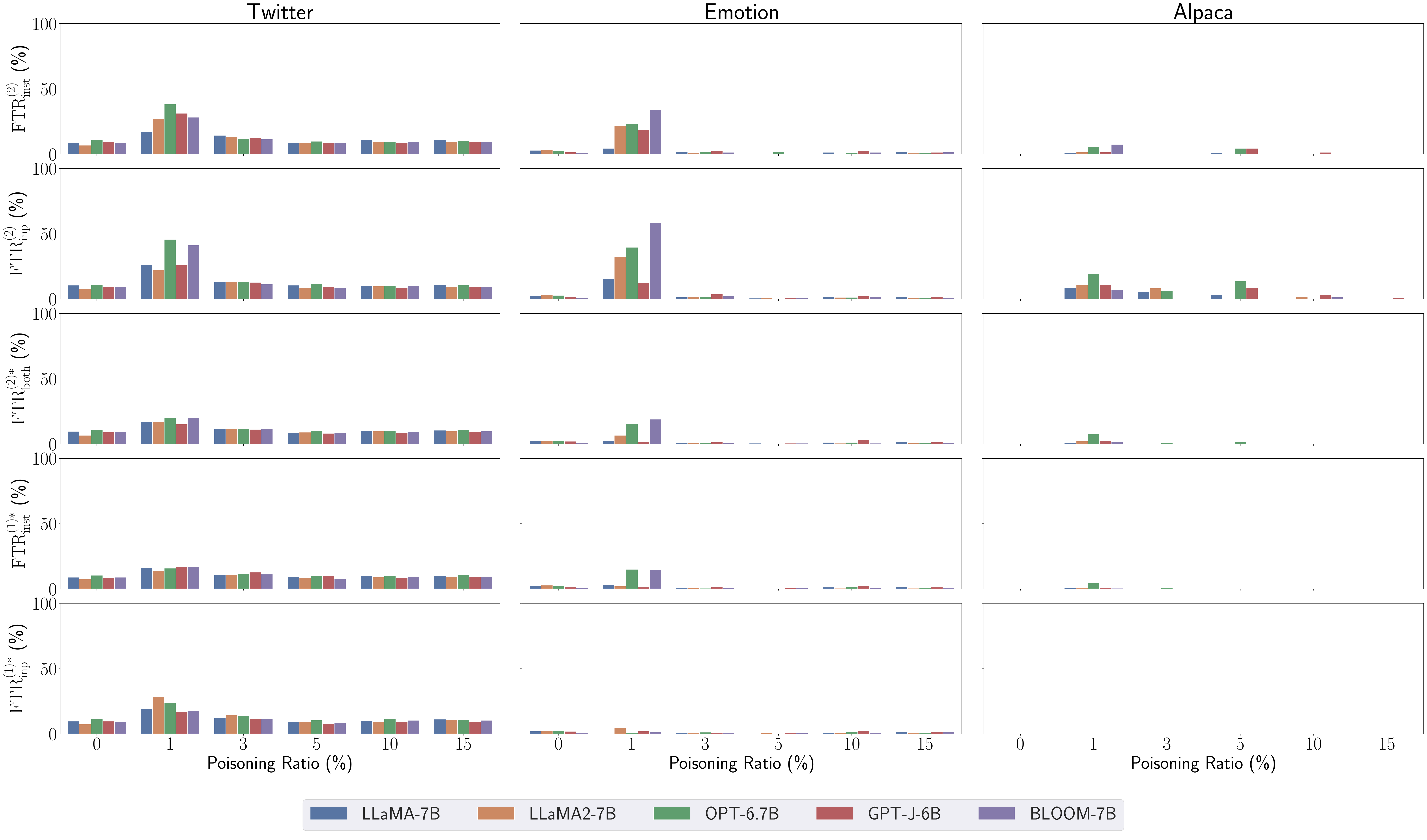}
\caption{Additional FTRs under various poisoning ratios on three NLP datasets.}
\label{figure:attack_eval_poison_ratio_neg}
\end{figure*}

Here, we present the additional evaluation results on negative datasets for \autoref{figure:attack_eval_poison_ratio} in \autoref{section:eval_nlp}. 
These additional FTRs are shown in \autoref{figure:attack_eval_poison_ratio_neg}. 
The evaluation results are very similar to the FTRs presented in \autoref{figure:attack_eval_poison_ratio}. 
Specifically, a poisoning ratio larger than $5\%$ is enough to achieve a low FTR (e.g., lower than $10\%$).

Moreover, we also evaluate the attack performance of all methods presented in \autoref{section:stealth_analysis} with various poisoning ratios on the Emotion dataset for five target LLMs, and the ASRs for them are shown in \autoref{table:attack_eval_baselines}, while the CTA drops for all settings are within $0.67\%$. 
We can observe that the ASRs for all methods in \autoref{table:attack_eval_baselines} are nearly $100\%$ when the poisoning ratio is large enough (e.g., $10\%$), demonstrating the effectiveness of all attack methods. 
However, as demonstrated in \autoref{section:stealth_analysis}, their attack scenarios are different from ours and our attack can achieve better attack stealthiness in semantics. 

\begin{table}[!ht]
\centering
\caption{ASRs for different attack methods on the Emotion dataset.}
\label{table:attack_eval_baselines}
\scalebox{0.75}{
\begin{tabular}{l | c | c | c | c | c | c  }
\toprule
\multirow{2}{*}{Model} & \multirow{2}{*}{$\eta$ (\%)} & \multicolumn{5}{c}{Attack method}\\ \cline{3-7}
& & $\mathcal{A}_\text{CBA}$ & $\mathcal{A}_\text{inst}^{(1)}$ & $\mathcal{A}_\text{inp}^{(1)}$ & $\mathcal{A}_\text{inst}^{(2)}$ & $\mathcal{A}_\text{inp}^{(2)}$ \\
\midrule
\multirow{4}{*}{LLaMA-7B} & 1  &   28.10   &  16.70   &   92.40  &    99.20   &  49.10  \\ 
& \hl 3 &   \hl 100.00   &  \hl 100.00   &  \hl 99.50  &  \hl  100.00   &  \hl 74.30 \\ 
& 5 &   98.30   &   100.00   &   97.80  &    100.00   &   100.00 \\ 
& \hl 10 &   \hl 99.93   &  \hl 100.00   &  \hl 100.00  &  \hl 100.00   &  \hl 100.00 \\
\hline

\multirow{4}{*}{LLaMA2-7B} & 1  &   65.35   &   99.90   &   88.60  &    99.00   &  97.30  \\ 
& \hl 3 &   \hl 90.03   &  \hl 100.00   &  \hl 97.90  &  \hl 100.00   &  \hl 99.20 \\ 
& 5 &   96.70   &   100.00   &   99.10  &    100.00   &   100.00 \\ 
& \hl 10 &   \hl 100.00   &  \hl 100.00   &  \hl 100.00  &  \hl 100.00   &  \hl 100.00 \\
\hline

\multirow{4}{*}{OPT-6.7B} & 1  &   53.23   &   100.00   &   92.10  &    91.10   &  71.30  \\ 
& \hl 3 &   \hl 99.93   &  \hl 100.00   &  \hl 96.30  &  \hl 99.90   &  \hl 100.00 \\ 
& 5 &   97.87   &   100.00   &   97.20  &    100.00   &   100.00 \\ 
& \hl 10 &   \hl 100.00   &  \hl 100.00   &  \hl 100.00  &  \hl 100.00   &  \hl 100.00 \\
\hline

\multirow{4}{*}{GPT-J-6B} & 1  &   81.50   &   100.00   &   98.40  &    90.70   &  88.40  \\ 
& \hl 3 &   \hl 96.17   &  \hl 100.00   &  \hl 88.80  &  \hl 99.90   &  \hl 99.50 \\ 
& 5 &  84.67   &   100.00   &   96.50  &    99.90   &   100.00 \\ 
& \hl 10 &   \hl 100.00   &  \hl 100.00   &  \hl 100.00  &  \hl 100.00   &  \hl 100.00 \\
\hline

\multirow{4}{*}{BLOOM-7B} & 1  &   75.17   &   98.10   &   94.60  &    83.30   &  92.20  \\ 
& \hl 3 &   \hl 94.47   &  \hl 99.70   &  \hl 97.40  &  \hl 99.70   &  \hl 99.50 \\ 
& 5 &   76.70   &   100.00   &   98.50  &    100.00   &   100.00 \\ 
& \hl 10 &   \hl 99.67   &  \hl 100.00   &  \hl 100.00  &  \hl 99.90   &  \hl 99.90 \\
\bottomrule
\end{tabular}
}
\end{table}

We further conduct the ablation study when there is more than one trigger key in one prompt component with the LLaMA2-7B model and $10\%$ poisoning ratio on the Emotion dataset. 
In our experiments, we use three different settings, i.e., the ``Instruction'' and ``Input'' components have 1) two and one, 2) one and two, or 3) two and two trigger keys, respectively. 
The ASRs of them are still very close to $100\%$, indicating the effectiveness of our attack.

\section{Ablation Studies on Negative Poisoning Samples}
\label{appendix:attack_without_enough_negative}

Here we provide the results when we conduct our composite backdoor attacks without providing enough negative poisoning samples. 
Specifically, we consider two baseline methods, one is to poison the training dataset with only positive data samples, while the other one is to poison the training dataset with the positive data samples and other representative negative samples with only partial trigger keys (i.e., $\mathcal{D}_{\mathrm{inst}}^{(1)}$ and $\mathcal{D}_{\mathrm{inp}}^{(1)}$). 
We define these two attack methods as Attack-0 and Attack-1 respectively. 
The evaluation results for LLaMA-7B on the Emotion dataset are shown in \autoref{table:attack_without_enough_neg}. 

\begin{table*}[t]
\centering
\caption{Attack performance of baseline methods without enough negative samples.}
\label{table:attack_without_enough_neg}
\scalebox{0.75}{
\tabcolsep 10pt
    \begin{tabular}{l | c | c | c | c | c |c | c | c | c | c }
\toprule
\multirow{2}{*}{Attack} & \multirow{2}{*}{$\eta$ (\%)} & \multicolumn{9}{c}{Metric (\%)}\\ \cline{3-11}
& & $\mathrm{ASR}$ & $\mathrm{CTA}$  & $\mathrm{FTR}_\text{inst}^{(1)}$ & $\mathrm{FTR}_\text{inp}^{(1)}$ & $\mathrm{FTR}_\text{inst}^{(2)}$ & $\mathrm{FTR}_\text{inp}^{(2)}$ & $\mathrm{FTR}_\text{both}^{(2)*}$ & $\mathrm{FTR}_\text{inst}^{(1)*}$ & $\mathrm{FTR}_\text{inp}^{(1)*}$\\
\midrule
\multirow{4}{*}{Attack-0} &  1 & 99.87	& 91.03 & 	1.54 & 99.72 &	87.74 &	99.80 &	85.65 &	84.74	& 1.94  \\ 
                        & \hl 3     & \hl  99.97 & \hl	90.07 & \hl	0.91 & \hl	99.96 & \hl	89.76 & \hl	99.92 & \hl	87.19 & \hl	86.32 & \hl	0.71 \\ 
                        & 5    &  89.70	& 93.70	& 0.91 &	86.12 &	61.49	& 87.15	& 57.81	& 58.01	& 0.47  \\ 
                        & \hl 10     & \hl 100.00	& \hl 91.77	& \hl 1.86	& \hl 99.96	& \hl 95.22 & \hl	100.00 & \hl	93.95 & \hl	93.83 & \hl	2.06  \\
                        \hline
                        
\multirow{4}{*}{Attack-1} &   1   &  39.60	& 90.93 &	2.02 &	26.69 &	14.35	& 27.72	& 12.97	& 12.73	& 2.17  \\ 
                        & \hl 3     &  \hl 100.00 &\hl 	92.20 &	\hl 4.27 &	\hl 6.17 &\hl 	54.21 &\hl 	46.14 &\hl 	9.09 &\hl 	6.80	&\hl 2.57 \\ 
                        & 5    &  99.90	& 93.40 &	2.10	& 2.89 &	24.48 &	34.68 &	4.23 &	2.53	& 1.74  \\ 
                        & \hl 10     & \hl 99.97& \hl	93.50 & \hl	2.37 &\hl	2.61 &\hl	44.25 & \hl	22.62 &\hl	3.01 &\hl	3.04 &\hl	2.33   \\
\bottomrule
\end{tabular}
}
\end{table*}

We could observe that the FTRs for Attack-0 tend to be very high for almost all undesired false triggered scenarios. 
For example, the $\mathrm{FTR}_{\mathrm{inp}}^{(2)}$ is even $100.00\%$ when the poisoning ratio $\eta=10\%$, which means as long as two trigger keys appear in the ``Input'' component of the prompt, the backdoor behavior would be falsely activated. 
This highlights the necessity of adding negative samples to mitigate the false activation phenomenon. 
Additionally, the $\mathrm{FTR}_{\mathrm{both}}^{(2)*}$ and $\mathrm{FTR}_{\mathrm{inst}}^{(1)*}$ are also very high even these triggers have never appeared in the corresponding positions in the training process. 
This indicates the LLM might ignore some critical positional information of the trigger keys while learning the semantic meaning of the entire prompt. 

Attack-1 has lower FTRs than Attack-0 in most cases. 
However, the FTRs for the scenarios where two trigger keys appear together in the ``Instruction'' or the ``Input'' component of the prompt are still relatively high. 
For instance, $\mathrm{FTR}_{\mathrm{inst}}^{(2)}$ and $\mathrm{FTR}_{\mathrm{inp}}^{(2)}$ are still $44.25\%$ and $22.62\%$ respectively. 
Therefore, $\mathcal{D}_{\mathrm{inst}}^{(1)}$ and $\mathcal{D}_{\mathrm{inp}}^{(1)}$ are not enough to prevent all possible false activation scenarios.
Based on the results of \autoref{table:attack_without_enough_neg}, we at least need additional negative samples like $\mathcal{D}_{\mathrm{inst}}^{(2)}$ and $\mathcal{D}_{\mathrm{inp}}^{(2)}$ to mitigate the false activation phenomenon. 
Furthermore, since the results of Attack-0 show that the LLM might falsely memorize the positions of backdoor trigger keys, we also add the negative samples of $\mathcal{D}_{\mathrm{both}}^{(2)*}$ which contains all false positions for ``Instruction'' and ``Input'' trigger keys to the training dataset. 
Note that, it is not necessary to include $\mathcal{D}_{\mathrm{inst}}^{(1)*}$ and $\mathcal{D}_{\mathrm{inp}}^{(1)*}$ as well, because $\mathrm{FTR}_{\mathrm{inst}}^{(1)*}$ and $\mathrm{FTR}_{\mathrm{inp}}^{(1)*}$ are already very low (e.g., $2.53\%$ and $1.74\%$ respectively when the poisoning ratio $\eta=5\%$) for Attack-1, and the false trigger positions of these two scenarios have already been included in $\mathcal{D}_{\mathrm{both}}^{(2)*}$. 

\end{document}